\documentclass[twocolumn]{aastex631}

\usepackage{gensymb}
\usepackage{emptypage}
\usepackage{amsmath}
\usepackage{float}
\usepackage{url}

\shorttitle{solar modulation}
\shortauthors{Jiang et al.}
\graphicspath{{./}{figures/}}

\begin{document}

\title{A New Scenario of Solar Modulation Model during the Polarity Reversing}

\correspondingauthor{Sujie Lin, Lili Yang}
\email{linsj6@mail.sysu.edu.cn, yanglli5@mail.sysu.edu.cn}

\author{Jieteng Jiang}
\affiliation{School of Physics and Astronomy, Sun Yat-sen University, Guangzhou 510275, People’s Republic of China}
\affiliation{CSST Science Center for Guangdong-Hong Kong-Macau Great Bay Area, Zhuhai 519082, China}

\author{Sujie Lin}
\affiliation{School of Physics and Astronomy, Sun Yat-sen University, Guangzhou 510275, People’s Republic of China}
\affiliation{CSST Science Center for Guangdong-Hong Kong-Macau Great Bay Area, Zhuhai 519082, China}

\author{Lili Yang}
\affiliation{School of Physics and Astronomy, Sun Yat-sen University, Guangzhou 510275, People’s Republic of China}
\affiliation{CSST Science Center for Guangdong-Hong Kong-Macau Great Bay Area, Zhuhai 519082, China}
\affiliation{Department of Physics, University of Johannesburg, PO Box 524, Auckland Park 2006, South Africa}

\begin{abstract}


When entering the heliosphere, Galactic Cosmic Rays (GCRs) will encounter the solar wind plasma, reducing their intensity. This solar modulation effect is strongly affected by the structure of the solar wind and the Heliospheric Magnetic Field (HMF). To address the effect during the solar maximum of cycle 24, we study the solar modulation under a scenario in which the weights for $A=\pm1$ are determined by the structure of HMF, and the traveling time of GCRs simulated by SOLARPROP is taken into account.
We then fit the cosmic ray proton data provided by AMS-02 and Voyager in the energy range $4\,\mathrm{MeV}\sim30\,\mathrm{GeV}$, and confirm that the modulation time lag in this model is about nine months, which is consistent with the previous studies.
This model incorporates a more realistic description of the polarity reversing and provides a more reliable estimation of the solar modulation effect during the maximum activity period.


\end{abstract}

\keywords{galactic cosmic ray, AMS-02, machine learning, solar modulation, proton}

\section{Introduction}\label{sec:intro}

Cosmic rays (CRs) have been widely studied for more than a hundred years, since the first discovery by Austrian-American physicist Victor Hess. They are charged, energetic nuclei coming from far beyond the solar system, and are believed to be originated from extreme phenomena in the universe. Specifically, galactic cosmic rays (GCRs) are particles accelerated to high energies from some powerful astronomical objects or magnetic fields in our Milky Way.

After crossing the heliopause (HP), the boundary of the solar system, the GCRs enter the heliosphere, collide with the solar wind moving outward, and are affected by the heliospheric magnetic field (HMF) \cite{parker1958dynamics}. It encompasses a few effects such as diffusion, drift, convection, and adiabatic energy changes (see reviews by e.g. \cite{heber2006cosmic, moraal2013cosmic, Cliver:2013kxa, kota2013theory, potgieter2013solar, engelbrecht2017toward}). As a result, their local interstellar spectrum (LIS) at the boundary is modulated, and this modulation effect varies for different types and energies of particles. 

The study of solar modulation is essential not only for comprehending the modulation process but also for promoting the relevant research.
For instance, the study of the transport model of GCR within the Galaxy~\cite{Yuan:2017ozr} and indirect search for dark matter with the anomalous CR antiproton flux~\cite{Lin:2019ljc} were hindered by the uncertainties in the LIS. For a better understanding of the modulation model, which could help us determine the LIS more accurately, more evidence and observation is required.

Fortunately, the highly precise GCR data was obtained in the last decade. For example, the Voyager 1 spacecraft, launched in 1977, provided proton data at a few MeV upon crossing heliopause in August 2012 \cite{stone2013voyager}. Additionally, the Alpha Magnetic Spectrometer (AMS-02) has offered precise measurements of protons across a broad energy range of 0.5 GeV to a few TeV near the Earth \cite{aguilar2018observation}. As particles with energies below 30 GeV are much more affected by the solar system, their LIS has to be obtained according to modulation models. 

Therefore to numerically describe the instant propagation of GCRs, Parker's equation, as a kind of the Fokker-Planck equation, have been widely applied over recent decades (e.g., \cite{fisk1971solar, gleeson1979effects, potgieter1985drift, jokipii1981effects}\cite{potgieter2000heliospheric}\cite{potgieter2014modulation}). With the input of physical heliosphere model, solar modulation parameters, and the LIS, the propagation of GCRs can be simulated with tools like SOLARPROP, allowing for the calculation of their energy spectrum at Earth. On the other hand, the track-back method can also be performed to get the LIS. In our study, we derive the proton LIS for energies from 0.5 GeV to 30 GeV in two periods of low solar activity, by interpolating the Voyager 1 data, fitting AMS-02 data and computing with 
 SOLARPORP \cite{wang2019time, wang2022solar}. 


At present, although some results of solar modulation can fit well with data during quiet solar epochs, it remains a challenge during solar maximum, because of the more complex coronal structure \cite{mccomas2001ulysses}, the behavior of the solar wind and HMF. Nonetheless, some research groups have made progresses for the maximum activity period. For example, \cite{song2021numerical} used five modulation parameters to fit the observed data, \cite{shen2021solar} employed a force-field approach to obtain the best-fit parameters, and \cite{fiandrini2021numerical} introduced a weight to linearly combine of the fluxes with two polarities. In this work, we redefined the weight and took into account the difference of particle energies to successfully obtain the best-fit parameters during solar maximum.

This paper is organized as follows. In Section \ref{model}, the heliosphere model and diffusion model are described in details. In Section \ref{method}, we analyze the active period on cycle 24 and apply alternative code scheme for better efficiency. In Section \ref{result}, the modeling results are presented. Also, the LIS of proton and the best-fit parameters from May 2011 to October 2016 are provided. A summary and conclusion are presented in Section \ref{conc}.

\section{Numerical Model}\label{model}
When the GCRs enter the solar system, they suffer from energy loss and direction change, which results in a reduction of their intensity. The propagation of these charged particles can be described by the transport equation, which was firstly given by Parker in 1965 \cite{PARKER19659} in the form of the Fokker-Planck equation (FPE) without sources
\begin{equation}
\begin{aligned}
    \frac{\partial f(\mathbf{r},\mathbf{p},t)}{\partial t} = &\nabla(\mathbf{K^S}\cdot\nabla f(\mathbf{r},\mathbf{p},t))
   + \frac{1}{3}(\nabla\cdot \mathbf{V_{SW}})\frac{\partial f(\mathbf{r},\mathbf{p},t)}{\partial ln~p} \\
    & -(\mathbf{V_{SW}}+\mathbf{V_D})\cdot\nabla f(\mathbf{r},\mathbf{p},t),
    \label{TPE}
\end{aligned}
\end{equation}
where $f(\mathbf{r},\mathbf{p},t)$, as a function of position $\mathbf{r}$, momentum $\mathbf{p}$, and temporal variable $t$, describes the dynamic phase-space distribution of GCRs. On the right side of Equation \ref{TPE}, there are three terms describing the CR transportation processes of diffusion, adiabatic energy loss, convection and drift in the heliosphere respectively. The physical quantities involved include diffusion coefficient $\mathbf{K^S}$, solar wind velocity $\mathbf{V_{SW}}$ and drift velocity $\mathbf{V_D}$. Here $\mathbf{V_{D}}$ includes gradient-curvature drift \cite{jokipii1977effects, jokipii1979effects} and the heliosphere current sheet (HCS) drift \cite{potgieter1985drift, burger1989calculation, hoeksema1992large} and diffusion velocity. 

To find the solution to FPE, the time-backward numerical method with stochastic differential equations (SDEs) has become popular. The pseudo particles are simulated from the moment they reach the Earth and traced backward until they reach the heliopause. \cite{yamada1998stochastic, zhang1999markov, kopp2012stochastic, kappl2016solarprop}. For a stochastic process driven by Wiener process, the SDEs describe the particle position $d\textbf{r}$ in the form of
\begin{equation}
   \mathrm{d}\textbf{r} = (\nabla\cdot\mathbf{K^S}-\mathbf{V})\mathrm{d}t+ \mathop{\sigma}\limits^{\leftrightarrow} \cdot \mathrm{d}\textbf{W},
   \label{SDE1}
\end{equation}
where $dt$ is time, $\mathbf{V} = \mathbf{V}_{\mathbf{SW}} + \mathbf{V_D}$ is the global velocity of the particles, 
$\mathop{\sigma}\limits^{\leftrightarrow}$ is a third-order matrix satisfying $\mathop{\sigma} \limits^{\leftrightarrow} \cdot \mathop{\sigma} \limits^{\leftrightarrow} = 2\mathbf{K^S}$, $\mathrm{d}\mathbf{W}$ is a Wiener process related to a standard normal distribution $N(0,1)$. The kinetic energy $\mathrm{d}T$ of a cosmic ray particle with mass $m$ in $\mathrm{d}t$ time interval can be preformed as
\begin{equation}
\mathrm{d}T=\frac{2\mathbf{\left|V_{SW}\right|}}{3\mathbf{\left|r\right|}}\frac{T^2+2Tm}{T+m}\mathrm{d}t,   \label{SDE2}
\end{equation}
here $m$ is the mass of particle. With the constructed numerical method above, we adopted the public code SOLARPROP \cite{kappl2016solarprop} to perform the particle simulation. Based on this framework, one can change the propagation model according to various presumptions on the physical quantities. In this work, we applied a 2D model to describe these quantities inside the heliosphere following Ref.~\cite{potgieter2014modulation}.

\subsection{Heliosphere Model}
Both the diffusion coefficient $\mathbf{K^S}$ and the drift velocity $\mathbf{V_D}$ depend on the HMF and solar wind.
Previously the large-scale HMF, embedded into the outward-flowing solar wind, was given by Parker as an Archimedean spiral field \cite{parker1958dynamics}. However, as the turbulence of footprint of HMF on the sun surface, 
the transverse perturbation of HMF near the sun would significantly enhance the average magnitude in the polar region \cite{jokipii1989polar}. In this work, we adopt the HMF model performed in Ref.~\cite{fichtner1996cosmic}, which takes the transverse perturbation into account by modifying the magnitude of the Archimedean spiral field. This modification is supported by the measurements of the magnetic field in the polar regions of the heliosphere by Ulysses \cite{balogh1995heliospheric}. The modified HMF model can be written in the form
\begin{equation}
\left\{
\arraycolsep=1.4pt\def\arraystretch{1.9}
\begin{array}{l}
\boldsymbol{B}=A~B_0\dfrac{r^2_0}{r^2}(\boldsymbol{e_r}+\zeta\boldsymbol{e_\theta}-\psi\boldsymbol{e_\varphi}) \\
\zeta=\dfrac{r\delta(\theta)}{r_\odot sin(\theta)}\\
\psi=\dfrac{\Omega(r-r_\odot)sin(\theta)}{V_{SW}}   
\end{array}
\right.
\label{mag}
\end{equation}
where $\Omega = 2.7\times10^{-6}\mathrm{rad/s}$ is the rotation angular velocity of the sun, $r_\odot = 3 \times 695500~\mathrm{km}$ is the radius of the corona, $V_{SW}$ is the velocity of the solar wind, $B_0$ is the HMF observed at the reference position $r_0$. Here $A$ is the polarity of the field and could only be 1 or $-1$, the N pole of HMF located in the northern solar hemisphere in the case $A=1$ and vice verse, and $\delta(\theta)$ is presumed to follow the expression \cite{fiandrini2021numerical}
\begin{equation}
\delta(\theta) = \left\{
\begin{array}{ll}
3\times 10^{-3} \sin(\theta), & 1.7^\circ < \theta < 178.3^\circ \\
8.7 \times 10^{-5}, &  \textrm{else}\\
\end{array}\right..
\label{delta_theta}
\end{equation}

The observation shows that the speed of the solar wind $\mathbf{V_{SW}}$ changes with radial and polar position during periods of minimum solar activity \cite{bame1992ulysses, heber2006cosmic}. Along the radial direction of the equatorial plane, the wind speed keeps constant at $430~\mathrm{km/s}$ until it reaches the termination shock (TS). It decreases to about $170~\mathrm{km/s}$ after across the TS and finally becomes zero or moves tail-ward in the inner heliosheath because of the barrier of the heliopause (HP) \cite{krimigis2011zero}. While along the polar direction, $\mathbf{V_{SW}}$ increases from about $430~\mathrm{km/s}$ to $800~\mathrm{km/s}$ in the high polar region, as observed by \cite{heber2006cosmic}. The solar wind speed was given by \cite{potgieter2014modulation},

\begin{equation}
\begin{aligned}
\mathbf{V_{SW}}(r,\theta) = & V_0(1.475 \mp 0.4 \tanh[6.8(\theta-\frac{\pi}{2}) \pm (\frac{15\pi}{180} + \alpha)])\\
& \times [\frac{s+1}{2s}-\frac{s-1}{2s} \tanh(\frac{r-r_{TS}}{L})]\boldsymbol{e_r} \label{SW}
\end{aligned}
\end{equation}

where $V_0 = 400~\mathrm{km/s}$, $\theta$ is the polar angle, the distance of termination shock $r_{TS}=90~\mathrm{AU}$, $s=2.5$ and $L=1.2~\mathrm{AU}$. And $\alpha$ is the tilt angle that describes the angle of the HCS. For the same $\theta$, the radial variation of Equation \ref{SW} is a constant while the polar variation changes from $430~\mathrm{km/s}$ near the equator to $800~\mathrm{km/s}$ in the polar region. The HMF strength around the earth $B_0$, polarity $A$, and the tilt angle $\alpha$ in Equation \ref{mag} and \ref{SW} can be obtained from the observation.

\subsection{Diffusion Model}
In general, the full diffusion tensor is expressed as  {$\mathbf{K}=\mathbf{K^S}+\mathbf{K^A}$}. It includes  symmetric diffusion tensor {$ \mathbf{K^S} $}, which is diagonal, and asymmetric diffusion tensor {$\mathbf{K^A}$} as following,

\begin{equation}
\begin{aligned}
\mathbf{K}  =\left[
\begin{array}{ccc}
  K_{r\perp}   & -K_A  &  0  \\
    K_A &  K_{\theta \perp}  &  0 \\
    0 & 0 & K_\parallel  \\
\end{array}
\right]
= &
 \underbrace{
\left[
\begin{array}{ccc}
K_{r\perp}&0&0\\
0 & K_{\theta \perp} &0 \\
0 & 0 & K_\parallel \\
\end{array}
\right]
}_{\mathbf{K^S}} \\
& +
\underbrace{
\left[
\begin{array}{ccc}
0& -K_A & 0\\
K_A & 0    &  0\\
0&0     & 0\\
\end{array}
\right]
}_{\mathbf{K^A}}
\end{aligned}
\end{equation}

The symmetric part describes the normal diffusion effect while the asymmetric part describes the drift effect. In the symmetric part, $K_{\parallel}$ is the diffusion component parallel to the direction of the magnetic field, and $K_{r \perp}$  and $K_{\theta \perp}$ are two perpendicular diffusion coefficients in the radial direction and the polar direction, respectively. A typical empirical expression for  $K_\parallel$ is given by Ref.~\cite{potgieter2014modulation} in the form of
\begin{equation}
K_\parallel=\left(K_{0}\right)\beta\left(\frac{B_0}{\left|\mathbf{B}\right|}\right)\left(\frac{R}{R_0}\right)^a\left(\frac{\left(\frac{R}{R_0}\right)^m+\left(\frac{R_k}{R_0}\right)^m}{1+\left(\frac{R_k}{R_0}\right)^m}\right)^\frac{b-a}{m},
\label{diffu}
\end{equation}
where $K_{0}$ is a constant with an order of $10^{23} \mathrm{cm^2 s^{-1}}$, $\beta=v/c$ is the speed of the particle in the nature unit, $B_0$ is the value of HMF detected around the Earth, $R=p/Z$ is the particle rigidity, the reference rigidity $R_0=1~\mathrm{GV}$, and $m=3.0$ guarantees the smoothness of the transition.
The indexes $a$ and $b$ determine the slope of the rigidity dependence below and above a rigidity with the value $R_k=3~\mathrm{GV}$, respectively.

Perpendicular diffusion term in the radial direction is presumed to be ~\cite{giacalone1999transport}
\begin{equation}
K_{r\perp}=0.02~K_\parallel,
\end{equation}
while the polar perpendicular diffusion term is given in Ref.~\cite{potgieter2000heliospheric, balogh2008galactic}
\begin{equation}
K_{\perp\theta}=0.02 K_\parallel f_{\perp\theta}.
\end{equation}
The factor $f_{\perp\theta}$ satisfies the expression
\begin{equation}
f_{\perp\theta}=A^+ \mp A^-\tanh[8(\theta_A-90^\circ\pm\theta_F)],
\label{f_theta}
\end{equation}
where $A^\pm = (d\pm1)/2$, $\theta_F=35^\circ$, and $\theta_A = 90^\circ - \left|90^\circ - \theta\right|$.
This means that $K_{\perp\theta}$ is enhanced towards the poles by a factor of $d$ with respect to the value of $K_\parallel$ in the equatorial regions of the heliosphere. The enhance factor $d$ is set to be 3.

Plugging the asymmetric part into the diffusion term $\nabla(\mathbf{K^A}\cdot\nabla f)$ would lead to a cross-product-like result in the form of $\nabla\times\mathbf{B}\cdot\nabla f$.
This term could describe the drift effect caused by the uneven magnetic field, thus it was written as the drift velocity in Equation \ref{TPE}.
Under the assumption of weak scattering and full drift process, the average drift velocity is related to the rigidity $R$ and the charge $q$ of particles, and the strength of magnetic field $B$~\cite{burger1985drift, burger1987inclusion}:
\begin{equation}
\left\langle \mathbf{V_D} \right\rangle=\nabla\times(\frac{qR\beta}{3B}\frac{\mathbf{B}}{B})).
\end{equation}

The drift velocity can be divided into two parts, gradient-curvature drift velocity $\mathbf{V_G}$ from the magnetic field and HCS drift velocity $\mathbf{V_{HCS}}$. The two drift velocities are expressed with two factors, $f(\theta)$ and $\zeta(R)$, given as \cite{potgieter1985drift, burger2000rigidity},

\begin{equation}
\left\{
\begin{array}{l}
\arraycolsep=1.4pt\def\arraystretch{1.9}
\mathbf{V_G}=f(\theta)\zeta(R)\cdot\nabla\times(\dfrac{qR\beta}{3B}\dfrac{\mathbf{B}}{B}) \\
\mathbf{V_{HCS}}=\zeta(R)\dfrac{qR\beta}{3B}\dfrac{\mathbf{B}}{B}\nabla\times f(\theta) \\
f(\theta)=\dfrac{1}{\alpha_h}\tan^{-1}[(1-\dfrac{2\theta}{\pi})\tan\alpha_h] \\
\zeta(R)=\dfrac{(R/R_A)^2}{1+(R/R_A)^2}
\end{array}\right..
\end{equation}

Here the cut-off value $R_A$ is fixed to be 0.5 GV according to \cite{fiandrini2021numerical}, $f(\theta)$ is a transition function, which models a wavy neutral sheet near the equator plane. And $\zeta(R)$ is a reduction function, which describes the change of drift velocity for different momenta of particles. The angle $\alpha_h$ equals to $arccos(\frac{\pi}{2c_h}-1)$, here $c_h=\frac{\pi}{2}-\frac{1}{\pi} sin(\alpha+\frac{2r_L}{r})$, $\alpha$ is tilt angle and $r_L$ depends on the maximum distance that particle can be away from the HCS.
 
In summary, the diffusion coefficient has been well established, except three parameters, $K_{0}$, indices $a$ and $b$, which are obtained from the analysis of the experiment data.

\section{Analysis and Calculation} \label{method}

\subsection{The Model Parameters}

\begin{figure}
    \includegraphics[width=\columnwidth]{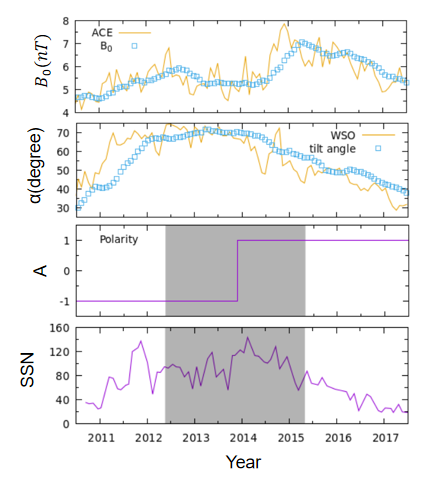}
    \caption{The observed data (solid line) of HMF $B_0$, tilt angle $\alpha$, polarity $A$ and sunspot number from the Advanced Composition Explorer (ACE) and the Wilcox Solar Observatory (WSO), separately, from 2011 to 2017. The square symbols are the average parameters of the ten periods of carrington rotation numbers. In the last two panels, the shaded regions represent the period from May 2012 to March 2015.}
    \label{fig:solar.png}
\end{figure}

To calculate the spectrum of GCRs near the Earth using the heliosphere model and the diffusion model, six parameters are needed, including three heliospheric parameters related to the solar system and three diffusion parameters. The heliospheric parameters are the strength $B_0$ of the HMF near the Earth, the tilt angle $\alpha$ of the HCS, and the polarity $A$ of the HMF, which can be obtained from observations, as shown in Figure \ref{fig:solar.png}. The value of $B_0$ is provided by the Advanced Composition Explorer (ACE), while the tilt angle and polarity are provided by the Wilcox Solar Observatory (WSO), represented by solid lines in the top two panels. The change in the magnetic field that is embedded in the solar wind typically takes about nine months to affect the motion of GCRs. This delay is referred to as the time lag \cite{Tomassetti:2017gkx, Orcinha_2019}. Considering that, we calculate the average field and tilt angle encountered by GCR particles during their journey from heliopause to Earth, as represented by the square symbol. The last panel in Figure \ref{fig:solar.png} shows the sunspot number (SSN) as a reference to compare the trend of $B_0$ and $\alpha$. It can be seen that $B_0$ and $\alpha$ increase with SSN, reaching maximum in February 2014, and the polarity reverses around this time. As for the other three diffusion parameters (normalization factor of diffusion $K_0$ and two spectral indices $a$ and $b$ in Equation 8), we can obtain them by fitting the observed data.

\subsection{Application of Machine Learning}

In this work, we applied the heliospheric model as described in Section \ref{model} and utilized SOLARPROP to simulate the propagation of GCR. In thi simulation, there are 30 energy bins from $0.4 \mathrm{GeV}$ to $30 \mathrm{GeV}$ and each bin has 2000 particles starting from the Earth. On average, it takes 1500 steps for each particle to reach the HP. Therefore a total of billions of steps are taken for all particles, and it needs about 10 minutes for SOLARPROP to complete one simulation. Running thousands of simulations can be quite time-consuming. To improve the efficiency, we employed a machine learning method, the LIBSVM library of Support Vector Machine (SVM) \cite{Chang:2011:LLS:1961189.1961199}, to replace the calculations of SOLARPROP. In order to construct the SVM model, we set a 5D parameter space with the following ranges:
\begin{itemize}
    \item $B_0$ in the range of $(3\sim8)\mathrm{nT}$
    \item $\alpha$ in the range of $15^\circ\sim75^\circ$
    \item $K_0$ in the range of $(0.001\sim1.5) \times 10^{23} \mathrm{cm^2 s^{-1}}$
    \item the indices $a$ and $b$ in the range of $0.001\sim3$
\end{itemize}
We randomly picked 40000 samples for $A=1$ and 50000 samples for $A=-1$ from this parameter space to train the SVM model. To ensure the reliability of the machine learning method, we performed detailed tests in Section \ref{4.1}.

\subsection{Analysis for Solar cycle 24}
In a recent study of the solar polar magnetic field during the  maximum activity in cycle 24, researchers found that the magnetic field underwent three reversals in the northern hemisphere (in May 2012, February 2014, and July 2014) and only one reversal in the southern hemisphere (in November 2013). This asymmetry of the magnetic field reversals has created a challenge in simulating solar modulation during this period, as the particles will experience magnetic fields with opposite directions. To address this issue, various methods have been proposed, such as adding more modulation parameters (as done by \cite{song2021numerical}) or simplifying the particle flux as a weighted sum of two spectra with different polarities (as proposed by \cite{fiandrini2021numerical}). We adopt the latter method and give the weight a physical meaning, as the ratio of the space occupied by the N-pole magnetic field in the heliosphere to the total space. Meanwhile, we also take into account the different propagation times of particles with different energies.

\section{Results} \label{result}
\subsection{The Local Interstellar Spectrum of proton} \label{4.1}
\begin{figure}
    \includegraphics[width=\columnwidth]{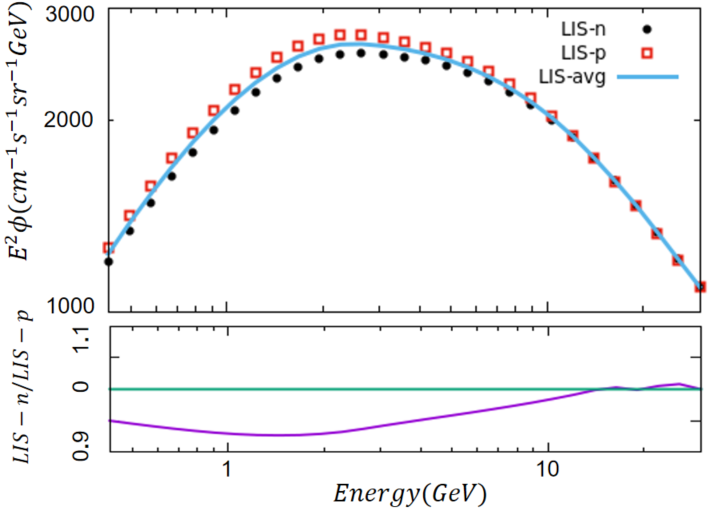}
    \caption{Derived the best-fit LIS of proton, constrained by the data from Voyager 1 and AMS-02. In the upper panel, LIS-n/LIS-p with negative/positive polarity are shown in dotted/solid lines, and the average LIS of LIS-n and LIS-p, labeled by LIS-avg. The ratio of LIS-n to LIS-p is shown in the second panel.}
    \label{fig:LIS.PNG}
\end{figure}

\begin{figure}
    \includegraphics[width=\columnwidth]{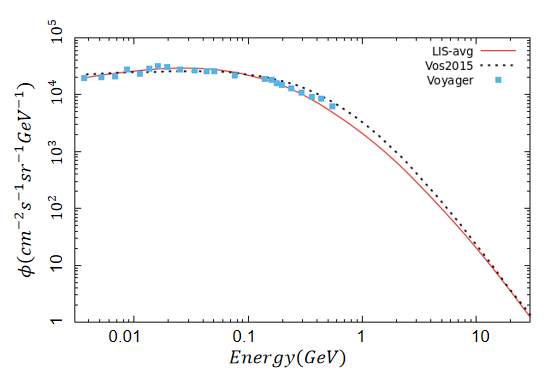}
    \caption{LIS-avg comparing with Voyager data in heliopause (blue squares) and the LIS from Reference~\cite{vosNEWMODELINGGALACTIC2015} (dotted line).}
    \label{fig:LIS-avg.PNG}
\end{figure}

\begin{figure*}
    \centering
    \includegraphics[width=0.7\textwidth]{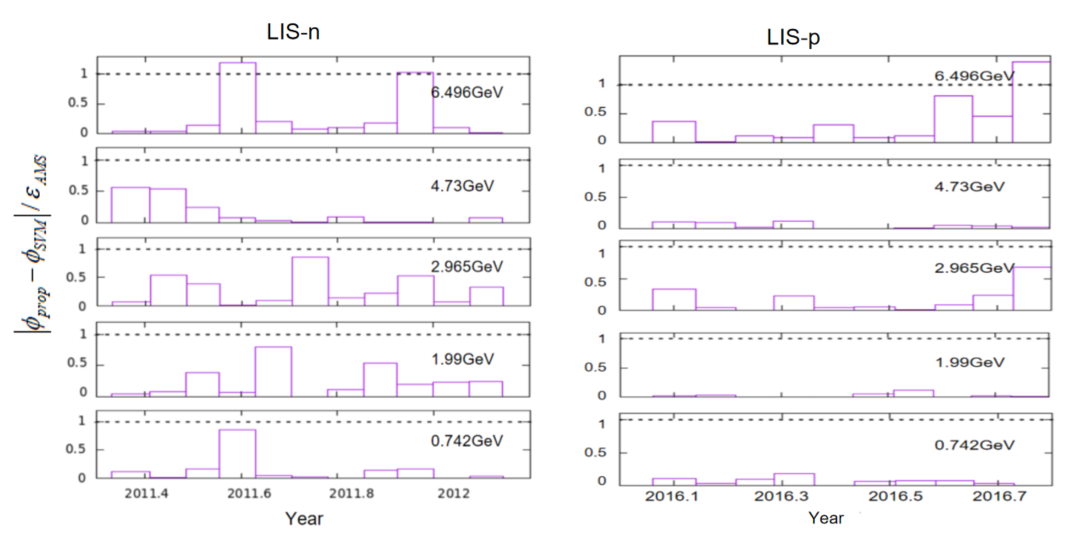}
    \caption{The time profile of the difference between $\phi_{prop}$ and $\phi_{svm}$, calculated from SOLARPRO and Libsvm, to the total err of AMS-02.}
    \label{fig:err.PNG}
\end{figure*}

The local interstellar spectrum (LIS) of protons represents the energy spectrum outside the heliopause. Voyager 1 crossed the heliopause in August 2012 and provided the LIS for protons at low energy (\textless~0.5 GeV). Additionally, energy spectra above a few GeV were measured by AMS-02 near the Earth. Solar modulation effects below 30 GeV are significant, but no directly observed LIS of protons has been obtained in the energy range between 0.4 GeV and 30 GeV. Therefore, the LIS in this range needs to be calculated.


We adopted both the data of Voyager 1 and AMS-02 to constrain the LIS. In order to avoid any unphysical wiggles occurring between these two datasets, we used the Akima spline~\cite{akimaNewMethodInterpolation1970} in the log-log plane to describe the LIS, with the reference energy listed at Table~\ref{tab:LIS}. The proton data observed by AMS-02 from two quiet periods were selected, which corresponded to Bartels' numbers 2426-2437 and 2470-2487, respectively, and corresponded to different HMF polarities. Two independent fittings were performed, and two proton LIS were obtained for these two periods as shown in Figure \ref{fig:LIS.PNG}, named LIS-n (dotted line) for negative polarity and LIS-p (solid line) for positive polarity.
The best-fit reference points for the two LIS result are listed at Table~\ref{tab:LIS}.
These two LIS are quite close to each other, with a relative difference of less than $10\%$.
It is well known that cosmic ray particles propagate in different paths for different HMF polarities due to the drift direction. A positive charge particle is likely to propagate inward along the heliospheric current sheet (HCS) in the negative polarity period, while it is likely to propagate along the polar regions in the positive polarity period. The reverse applies for the negative charge particle. Therefore, the difference between the two LIS shows the potential systematic error inside our configuration of HMF, HCS, and solar wind. As this difference is acceptable and do not affect our investigation of the maximum activity period, we left it for the future study. In this work, the two LIS were averaged to obtain a unified LIS (LIS-avg) for the following work, which is shown as square symbols in Figure \ref{fig:LIS.PNG}.
We also compared this LIS with the data of Voyager 1 and the LIS from Reference~\cite{vosNEWMODELINGGALACTIC2015} in Figure~\ref{fig:LIS-avg.PNG}, a difference less than $60\%$ was found.

\begin{table*}[htbp!]
    \centering
    \begin{tabular}{cccccccccc}
        \hline
        E[GeV] & $3.71\times10^{-3}$ & 0.426 & 2.64 & 14.0 & 16.3 & 19.0 & 22.1 & 25.8 & 30 \\
        \hline
        LIS-n[$\mathrm{GeV^{-1}cm^{-2}sr^{-1}}$] &  20060  & 6620 & 366 & 8.85 & 6.02 & 4.06 & 2.73 & 1.82 & 1.22 \\
        LIS-p[$\mathrm{GeV^{-1}cm^{-2}sr^{-1}}$] &  20060 & 6970 & 391 & 8.87 & 6.00 & 4.06 & 2.71 & 1.81 & 1.22 \\
        \hline
    \end{tabular}
    \caption{The reference points of the best-fit LIS with Akima spline interpolation.}
    \label{tab:LIS}
\end{table*}

To evaluate the validity of machine learning, SOLARPROP and LIBSVM were applied with given LIS-n and LIS-p, and the fluxes $\phi_{svm}$ and $\phi_{prop}$ were obtained after solar modulation using the best-fit parameters. By comparing the difference between $\phi_{prop}$ and $\phi_{svm}$ and the total error of AMS-02, the validity is given in Figure \ref{fig:err.PNG}. The ratio are mostly less than one for both LIS-n and LIS-p, which means the difference between $\phi_{prop}$ and $\phi_{svm}$ from the two methods is smaller than the total error and can be neglected in our analysis.

\subsection{Quiet Periods}
\begin{figure}
    \centering
    \includegraphics[width=\columnwidth]{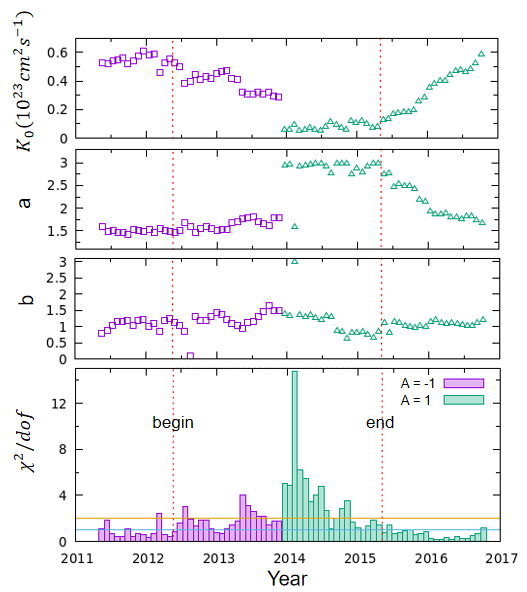}
    \caption{Results of the best-fit parameters, $K_0$, $a$ and $b$ for two periods, $A=-1$ (square) and $A=1$ (triangle), and the corresponding reduced chi-square in the last panel, here the dof equals to 34. The vertical dashed lines indicate the beginning and the end of the reversal epoch.}
    \label{fig:quiet.png}
\end{figure}

To determine the best-fit parameters for cycle 24 during solar maximum, we analyzed the full set of data of AMS-02 from May 2011 to October 2016. In this data set, we bin the data into 34 energy bins from 0.47 GeV to 24.71 GeV, namely 34 degree of freedom (Dof). The best values for the parameters $K_0$, $a$, and $b$ are presented in Figure \ref{fig:quiet.png}. Notably, two parameters, $K_0$ and index $a$, exhibit a sudden change in November 2013. Specifically, the diffusion coefficient $K_0$ decreases, while index $a$ increases. This change can be attributed to the polarity shift illustrated in the third panel of Figure \ref{fig:solar.png}. In contrast, the value of index $b$ remained stable throughout the analyzed period.

As shown in the last panel of Figure \ref{fig:quiet.png}, the value of reduced chi-square is less than 1 during two quiet periods. But the value increases to more than 1 and keeps on for a long time, the increasing rate is more than 60\%, and the time points are May 2012 and May 2015. So we conclude that the solar magnetic field reversal occurred from May 2012 to May 2015. Some other works hold the same point of view~\cite{pishkalo2016dynamics, gopalswamy2016unusual} through the observed polar magnetic field on the surface of the Sun.

The last panel of Figure \ref{fig:quiet.png} shows that the fitting result from May 2012 to May 2015 is unsatisfactory, particularly after late 2013 (with $\chi^2 > 2$).
The reason for this could be attributed to the maximum solar activity during this period, as well as the polarity reversal of the HMF.
Given the complexity of this situation, a more sophisticated model is necessary.

\subsection{Maximum Activity} \label{act}

During periods of maximum activity, the sign of the large-scale magnetic field can vary at different positions, even within the same hemisphere. As a result, cosmic rays will encounter the HMF with different polarities along their path. Ideally, the magnetic field at the location of each cosmic ray should be simulated, but currently, it is not possible to detect the magnetic polarity and path of every cosmic ray within the heliosphere. Nonetheless, some progress has been made in simplifying this process. For example, \cite{fiandrini2021numerical} introduced a weight term, denoted by $P$, to calculate the final spectrum, $\phi_f$ near the Earth. This spectrum is the weighted sum of the spectra with two polarities, $\phi^-(E)$ with $A=-1$, and $\phi^+(E)$ with $A=1$.

\begin{equation}
\phi_f(E)=\phi^- (E)(1-P)+\phi^+ (E)P \label{weight}
\end{equation}

\begin{figure}
    \centering
    \includegraphics[width=\columnwidth]{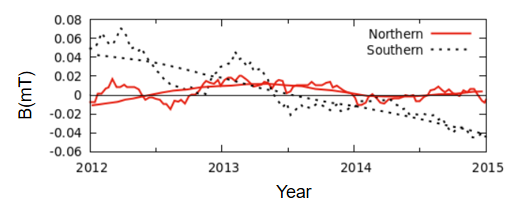}
    \caption{The two polar magnetic fields (above $55^\circ$), northern pole (solid line) and southern pole (dotted line). The wave line is 10 days averaged, Data is from the Wilcox Solar Observatory (\cite{NBS_2021}).}
    \label{fig:Pmag.png}
\end{figure}

\begin{figure}
    \centering
    \includegraphics[width=\columnwidth]{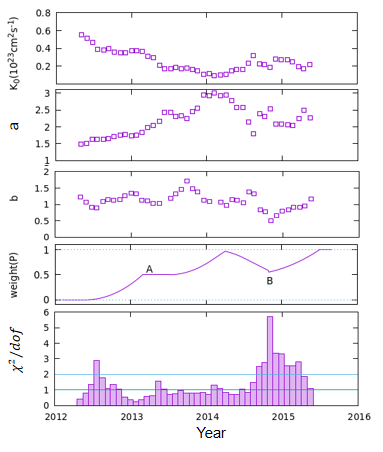}
    \caption{The best-fit parameters during maximum activity from May 2012 to May 2015. The final flux near the Earth is the weighted sum of the spectra with two polarities. The weight $P$ is the ratio of the space occupied by the N-pole magnetic field in the heliosphere to the total space, shown in the forth panel. The reduced $\chi^2$ is in the last panel.}
    \label{fig:act2.png}
\end{figure}

Here we employ a similar approach. Considering that magnetic field transports at solar wind speeds, we define the weight as the ratio of the space occupied by the N-pole magnetic field in the heliosphere to the total space. Figure \ref{fig:Pmag.png} shows that the directions of the polar magnetic field (above $55^\circ$) have changed over time for both the northern and southern hemispheres. The northern hemisphere experienced magnetic field reversals three times in May 2012, February 2014, and August 2014, while the southern hemisphere experienced one in July 2013. The calculated weight is presented in the forth panel of Figure \ref{fig:act2.png}, where two structures, platform A and valley B, are evident. These structures can be explained by the temporary stability of the solar field in the first half of 2013 and the change to a negative field in the northern hemisphere in the first half of 2014. Using the specified weight, the best-fit parameters are shown in the upper three panels in Figure \ref{fig:act2.png}, where the parameters change continuously. The parameter $K_0$ gets at a minimum in February 2014, and the index $a$ reaches a maximum at the same time. Compared to the SSN change in Figure \ref{fig:solar.png}, which also reaches an extreme value in February 2014, the parameters show obvious trends with the change in solar activity. However, although these three parameters show clear trends, the reduced $\chi^2$ in the last panel of Figure \ref{fig:act2.png} still has a high value, especially from August 2014 to April 2015, which makes the results somewhat unsatisfactory. Thus, further improvement is necessary.

\begin{figure}
    \centering
    \includegraphics[width=\columnwidth]{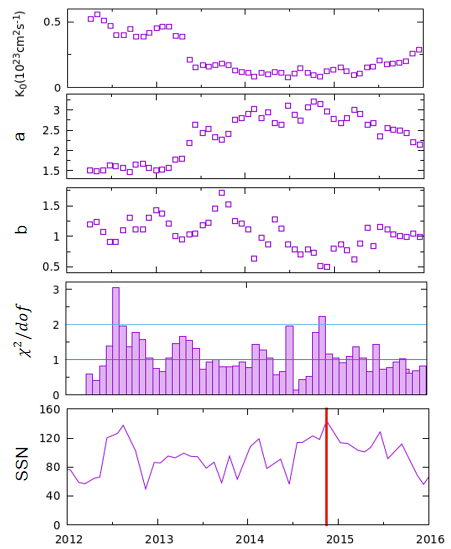}
    \caption{The best-fit parameters during maximum activity considering different traveling times of GCRs with different energies in the first three panel. The reduced $\chi^2$ is the forth panel. The panel of SSN($T-\Delta T_{lag}$) is the fifth panel, here $\Delta T_{lag}$ is time lag, equals to 9 months.}
    \label{fig:act3.png}
\end{figure}

To improve our results, we take into account the different traveling times of GCRs with different energies. To achieve this, we have utilized the time data simulated by SOLARPROP as a reference, which provides the proton traveling time for each energy level. We have used the average values of parameters such as $K_0=0.3\times10^{23}cm^{-2}s^{-1}sr^{-1}GeV^{-1}$, $a=1.80$, $b=0.989$, $B_0=5nT$, $\alpha=70^\circ$, and $A=1$. These values have been selected based on the points with $\chi^2/dof$ greater than 1 in the last panel of Figure \ref{fig:act2.png}. The time range for each energy bin is between 1.87 and 163 days, and particle traveling time decreases as their energy increases. By adding this information, the best-fit parameters and reduced $\chi^2$ are shown in Figure \ref{fig:act3.png}. The three diffusion parameters have the same trends as shown in Figure \ref{fig:act2.png}, but the extreme values are in November 2014, and the index $b$ is stable as always. Compared with the time reaching extreme values in Figure \ref{fig:act2.png}, the difference of time is nearly 9 months, which is just a time lag. The reduced $\chi^2$ in the last panel of Figure \ref{fig:act3.png} shows that 60\% of them have a value less than 1, 90\% less than 2, and only one has a maximum of 3.7. Therefore, we conclude that these best-fit parameters are reliable.

\section{Conclusion} \label{conc}

This study examines the solar modulation of Galactic Cosmic Rays (GCRs) and presents a new Local Interstellar Spectrum (LIS) of protons during solar activity in cycle 24, as seen in Figure \ref{fig:LIS-avg.PNG}. The final spectrum near the Earth during the period of solar maximum is obtained with a weight in Equation \ref{weight}, which is defined as the ratio of the space occupied by the N-pole magnetic field in the heliosphere. The weight is used to fit the final spectrum, which is equal to a weighted sum of two spectra with both polarities. The best-fit diffusion parameters are then determined, and their trends are shown in Figure \ref{fig:act2.png}. The normalization diffusion coefficient $K_0$ reached a minimum in February 2014, while index $a$ reached its maximum at the same time. In contrast, index $b$ does not exhibit a regular change. However, due to the different motion times of particles with different energies in space, the ratio of the magnetic field occupying needs to be modified for each energy bin. The modified best-fit parameters are shown in Figure \ref{fig:act3.png}, which also have one extreme point, but this time it occurs in November 2014, which is nine months later than the previous one. This delay time represents the time lag for the solar magnetic field to affect the energy spectrum.

The time lag was discussed with different methods in the literature. For example, in Ref.~\cite{fiandrini2021numerical}, they established a relationship between the parameters and the sunspot number (SSN) at the epoch $t-\Delta T_{lag}$, and found that the curve of $K_0$ vs SSN approaches a single-valued function. The $\Delta T_{lag}$ finally given by this method is about 11 months, which is comparable with the 9 months in our study.

To improve the reliability of LIS in future work, there are two main steps that should be taken. Firstly, it is important to overlap the energy range between Voyager data and that near the earth. Currently, the AMS-02 data is used, but there is no overlapped energy range between the Voyager and AMS-02 data. To address this issue, data from PAMELA can be utilized, the energy range is ($0.088\sim46.5$GeV), which covers the energy range of interest. However, PAMELA data is only available during negative polarity, which limits the ability to obtain the LIS with positive polarity. Therefore, it is necessary to wait for PAMELA to release new observations. Secondly, to account for the difference in particle motion time, it is important to consider the motion time of particles corresponding to diffusion param  eters rather than the average parameters. By doing so, the reliability of the results can be improved. In summary, the two next steps to improve the reliability of LIS are to utilize PAMELA data and to consider the motion time of particles corresponding to diffusion parameters.

\section{Acknowledgements}
 We thank E. Fiandrini and N. Tomassett for the valuable discussion. This work is supported by the National Natural Science Foundation of China (NSFC) grants 12205388, 12005313, 42150105, and 12261141691.

\vspace{5mm}

\bibliography{ref}
\bibliographystyle{aasjournal}

\end{document}